# β-Ga$_2$O$_3$-Based Radiation Detector for Proton Therapy

Hunter D. Ellis[1], Imteaz Rahaman[1], Apostoli Hillas[1], Botong Li[1], Vikren Sarkar[2], and Kai Fu[1, a]

[1]*Department of Electrical and Computer Engineering, The University of Utah, Salt Lake City, UT 84112, USA*

[2]*Department of Radiation Oncology, The University of Utah, Salt Lake City, UT 84112, USA*

**Abstract**

Intensity-modulated proton therapy (IMPT) is an advanced cancer treatment modality that offers significant advantages over conventional X-ray therapies, particularly in its ability to minimize radiation dose beyond the tumor target. This reduction in unnecessary irradiation exposure significantly lowers the risk to surrounding healthy tissue and reduces side effects compared to conventional X-ray treatments. However, due to the high complexity of intensity-modulated proton therapy plans, each plan must be independently validated to ensure the safety and efficacy of the radiation exposure to the patient. While ion chambers are currently used for this purpose, their limitations—particularly in angled beam measurements and multi-depth assessments—hinder their effectiveness. Silicon-based detectors, commonly used in X-ray therapy, are unsuitable for IMPT due to their rapid degradation under proton irradiation. In this study, a β-Ga$_2$O$_3$-based metal–semiconductor–metal (MSM) detector was evaluated and compared with a commercial ion chamber using a MEVION S250i proton accelerator. The β-Ga$_2$O$_3$ detector demonstrated reliable detection of single-pulse proton doses as low as 0.26 MU and exhibited a linear charge-to-dose relationship across a wide range of irradiation conditions. Furthermore, its measurement variability was comparable to that of the ion chamber, with improved sensitivity observed at higher bias voltages. These results highlight the strong potential of β-Ga$_2$O$_3$ as a radiation-hard detector material for accurate dose verification in IMPT.

[a] Author to whom correspondence should be addressed. Electronic mail: kai.fu@utah.edu

**Introduction**

Cancer remains a significant global health challenge, with high mortality rates every year. In the quest for more effective treatments, intensity-modulated proton therapy (IMPT) has emerged as a promising alternative to conventional X-ray therapy. IMPT is advantageous due to its utilization of the spread-out Bragg peak, which enables the uniform delivery of high-energy doses at a designated depth within the tumor while minimizing the exit dose, sparing non-cancerous tissues from unnecessary irradiation, especially for tumors located near radiosensitive organs (Fig. 1)[1, 2]. Studies have shown that IMPT can reduce the risk of secondary cancers by more than 60% compared to traditional X-ray radiation therapy[3, 4]. In addition, IMPT reduces side effects associated with treatment, improving patient outcomes and quality of life[5-8].

Despite these benefits, the implementation of IMPT requires meticulous planning and validation for each individual treatment session[9, 10]. Currently, ion chambers are used for the validation process; however, their utility is limited by their inability to perform angled beam testing, which restricts their effectiveness in this application[11]. Angled beam deliveries are essential for patient care, and they are accomplished through the movement of the gantry, as shown in Fig. 1. Traditional silicon-based detector arrays, which are capable of angled beam testing and are used in X-ray therapy, cannot be used for IMPT due to their rapid degradation under proton irradiation.

To address this limitation, β-Ga$_2$O$_3$, an ultra-wideband gap semiconductor, has emerged as a promising alternative. β-Ga$_2$O$_3$ is characterized by its considerable bond strength, which confers excellent resistance to radiation damage[12, 13]. The strength of this bond leads to a large bandgap, and a high displacement energy ($E_d$). A high $E_d$ is crucial for proton radiation detectors, as it indicates the amount of energy required to remove an atom from the lattice and form a defect. Figure 2(a) illustrates the relationship between displacement energy and the bandgap for various common semiconductors[13, 14]. β-Ga$_2$O$_3$ has a relatively high displacement energy compared to Si and many other semiconductor materials, resulting in relatively higher radiation resistance.

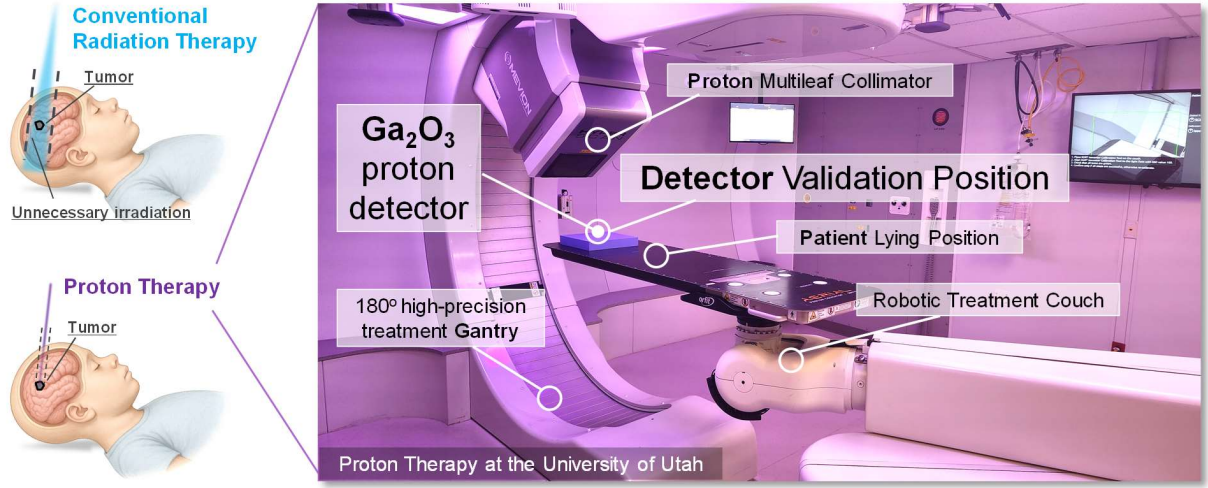

**Fig. 1**. Schematic comparison between conventional radiation therapy and proton therapy, alongside a photograph of the proton radiation treatment room in this study.

Another key parameter for materials used for radiation detectors is the electronic stopping power ($S_e$), which quantifies the energy transferred to electrons in the lattice as radiation passes through the crystal per unit length, and it is a primary method of kinetic energy transfer between the radiation and the detector material at the high energies seen in IMPT[15, 16]. This is particularly important for IMPT since most of the proton radiation travels completely through the device, meaning only a fraction of the total energy is deposited in the detector. As the $S_e$ increases, the deposited energy also increases. Figure 2(b) presents the relationship between $S_e$ and the atomic mass $Z$ for common semiconductors. The $S_e$ values for each material was calculated using a SR-NIEL calculator for proton radiation at 200 MeV[17]. The high electronic stopping power of β-$Ga_2O_3$, allows for increased responsivity of detectors made from this material.

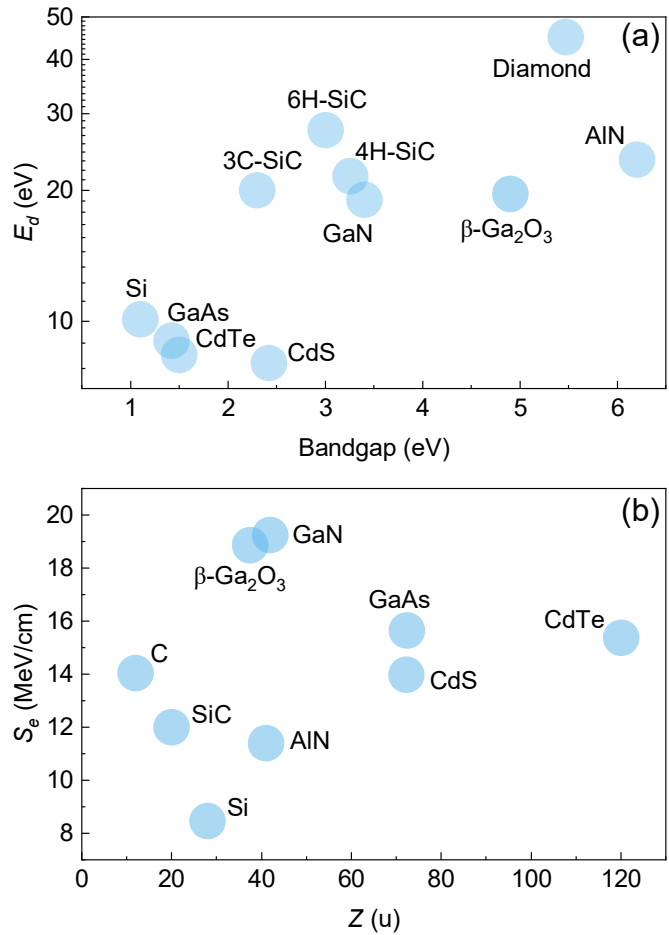

**Fig. 2**. Radiation hardness of different semiconductors. (a) Displacement energy ($E_d$) of different common semiconductor materials. (b) Electronic stopping power ($S_e$) of different common semiconductors.

β-Ga$_2$O$_3$ based radiation detectors have been successfully utilized for the detection of many varieties of radiation, including alpha particles, neutrons, and X-rays[18-26], showing robust performance. The high energy radiation tolerance of β-Ga$_2$O$_3$ based devices have also been demonstrated with UV photodetectors continuing to function with a responsivity of 25 A/W after exposure to proton doses of $10^{15}$ cm$^{-3}$ at 5 MeV, along with a MOSFET that had a gamma ray tolerance of 1.5 MGy[27, 28]. The radiation resilience of β-Ga$_2$O$_3$ diodes has also been demonstrated with NiO/β-Ga$_2$O$_3$ pn-diode receiving a proton radiation fluence of $2\times10^{13}$ cm$^{-2}$ while still maintaining a high forward current density of 137 A/cm$^2$, and β-Ga$_2$O$_3$ Schottky barrier diodes being exposed to a proton fluence of $5\times10^{12}$ cm$^{-2}$ and still achieving a forward current density of 190 A/cm$^2$ respectivly[29, 30]. The carrier removal rate from proton radiation in β-Ga$_2$O$_3$ is also lower compared to other wide bandgap materials, such as 4H-SiC, allowing for higher total radiation exposure without severe deterioration[31]. These results suggest that β-Ga$_2$O$_3$ is highly resilient to high-energy proton radiation, making β-Ga$_2$O$_3$ an ideal material for proton radiation detection. β-Ga$_2$O$_3$ radiation detectors have also been shown to have a high internal gain due to extremely low hole mobility, which is also seen in other semiconductors such as GaN[32-35]. However, despite its promising radiation tolerance, the behavior and performance of β-Ga$_2$O$_3$ under proton irradiation, particularly in the context of proton therapy, have received limited investigation.

In this work, the feasibility of using β-Ga$_2$O$_3$ detectors for IMPT was evaluated. Key performance data are presented, including the transient current response of the detector to a single proton radiation pulse, multiple pulses, and a raster scan of proton pulses. The charge response of the β-Ga$_2$O$_3$ detector was found to be linear across the full dose range of the MEVION S250i accelerator, from minimum to maximum output. Additionally, the detector achieved a higher sensitivity than a standard ion chamber for the same proton dose. The variability in repeated measurements with the β-Ga₂O₃ detector was comparable to that observed with the ion chamber, demonstrating consistent performance.

**Methods**

The β-Ga$_2$O$_3$-based metal-semiconductor-metal (MSM) proton detectors (GOPD) were fabricated according to the design illustrated in Fig. 3(a). An optical microscope image of the detector is also shown in Fig. 3(a). The design comprises metallic fingers, each 10 μm thick, with a 10 μm spacing between adjacent fingers. A detector with 25 fingers was manufactured, with a finger length of 490 μm. The β-Ga$_2$O$_3$ used was an unintentionally doped sample grown on sapphire. The β-Ga$_2$O$_3$ film was epitaxially grown by an Agnitron MOCVD system, which lasted 90 minutes at a temperature of 840 ℃. The Ar shroud gas flow rate was 500 sccm, with O$_2$ at 800 sccm, Ar at 1100 sccm, and the TEG precursor at 130 sccm. The chamber pressure was maintained at 60 Torr.

For device fabrication, a Ni (60 nm)/Au (100 nm) bilayer was deposited onto the β-Ga$_2$O$_3$ surface using an e-beam, followed by a lift-off process to define the detector structures. The wafer was then affixed to a printed circuit board using PELCO conductive silver paint. Subsequently, wire bonding was performed to connect the detector to the PCB pins, and copper wires were soldered to ensure stable electrical connections. The PCB was mounted on a fixture to provide stability during the measurements.

The device operates by generating free charge carriers within the active region of the GOPD through interactions between the incident proton radiation and the β-Ga$_2$O$_3$ material. When a bias voltage is applied across the electrodes, it modulates the Schottky barriers at the contacts, reducing one barrier while increasing the other. Under this biased condition, the proton-generated electrons are driven away from the electrode with the higher barrier and are able to surmount the lower barrier, thereby contributing to a measurable current, as illustrated in Fig. 3(b).

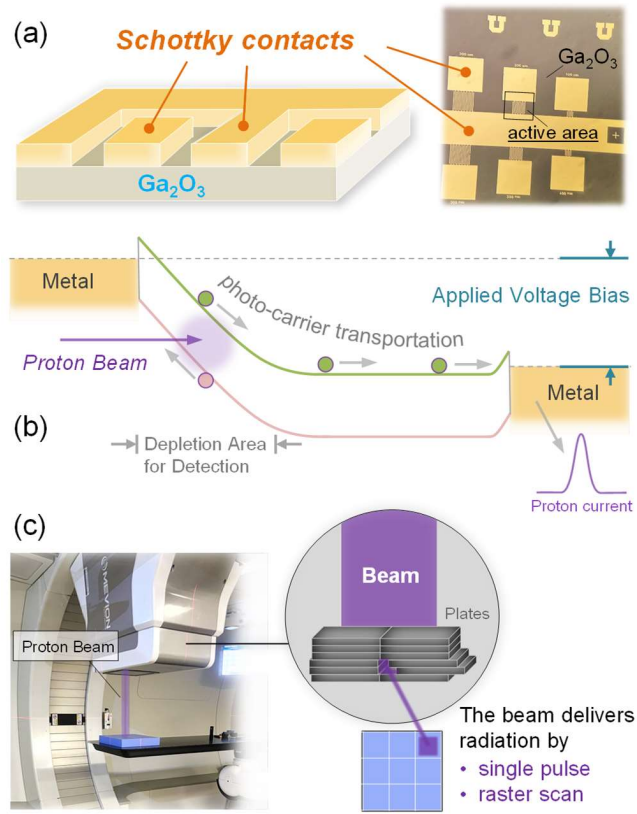

**Fig. 3**. Detector structure and detection mechanism. (a) Diagram along with an optical microscope image of the GOPD. (b) Band diagram of an MSM detector with an applied bias and proton radiation creating free carriers. (c) Placement of the GOPD relative to the proton emitter, along with visualization of the single-pulse and raster-scan proton delivery methods.

The test board and GOPD were positioned on a carbon fiber patient bed directly beneath the proton beam emitted by the MEVION S250i accelerator, with its placement relative to the machine shown in Fig. 3(c). Once accelerated to their therapeutic energy, the protons are directed through a collimation device equipped with dynamically adjustable metal plates, which move back and forth to shape the beam opening, allowing the finely focused pencil beam to conform precisely to the three-dimensional (3D) contours of the tumor. Radiation is delivered layer by layer throughout the tumor volume, in a manner like 3D printing. To facilitate data collection while avoiding exposure to the proton beam, cables were used to connect the GOPD on PCB board to a Keithley 2470 SourceMeter located outside the proton treatment room. A PPC05 plane parallel chamber was placed below the detector to obtain ion chamber measurements of the proton radiation to compare with the GOPD.

In this study, two beam delivery modes are characterized: single-pulse mode at 189.9 MeV and raster-scanning mode at 150 MeV. In single-pulse mode, a high-intensity proton burst is delivered in a short duration to a fixed location. In contrast, raster-scanning mode sweeps the pencil beam across the target area in a controlled pattern, enabling spatial dose modulation with high precision, which is often used to treat patients. The raster-scan method scans the emitter across the β-$Ga_2O_3$ encompassing a 25 $cm^2$ square area, where 441 spots, equally spaced, in the 25 $cm^2$ area will be irradiated with a small-time delay between each spot. Fig. 3(c) shows an illustration of the pulse shaping method through plates in the MEVION S250i accelerator, along with a representation of the single-pulse measurement and the raster-scan area.

**Results**

Single-pulse measurements were performed with the detector operating at various bias voltages. Figure 4(a) illustrates how the transient current response of the GOPD varies with applied bias in the range of 4 V to 8 V from the minimum proton radiation dose level of the MEVION S250i accelerator of 0.26 MU to the maximum of 40 MU. Figure 4(a) shows that the GOPD, when biased at 8 V, successfully detects proton doses ranging from 0.26 MU to 40 MU. At lower biases (4 V and 6 V), the detection threshold increases, indicating reduced sensitivity at low doses. This indicates that higher bias enhances the GOPD's sensitivity, enabling detection of the lowest dose levels from the accelerator—a critical requirement for accurate measurement of minimal proton pulses. In contrast, the GOPD biased at 6 V and 4 V was unable to detect doses below 1 MU and 2 MU, respectively. The observed increase in sensitivity with applied voltage is attributed to enhanced band bending in the active region of the MSM detector, resulting in a larger depletion volume, more efficient carrier collection, and increased generation of electron-hole pairs[36]. The current pulses were fitted using the following formula:

$$I = I_0 + A \times (1 - \exp(\frac{-(t-t_0)}{\tau_{rise}})) \times \exp(\frac{-(t-t_0)}{\tau_{decay}}) \quad (1)$$

where $I$ is current, $t$ is time, $\tau_{rise}$ is the rise time constant, $\tau_{decay}$ is the decay time constant, $t_0$ is the delay time of the pulse, $A$ is a constant, and $I_0$ is the leakage current[37].

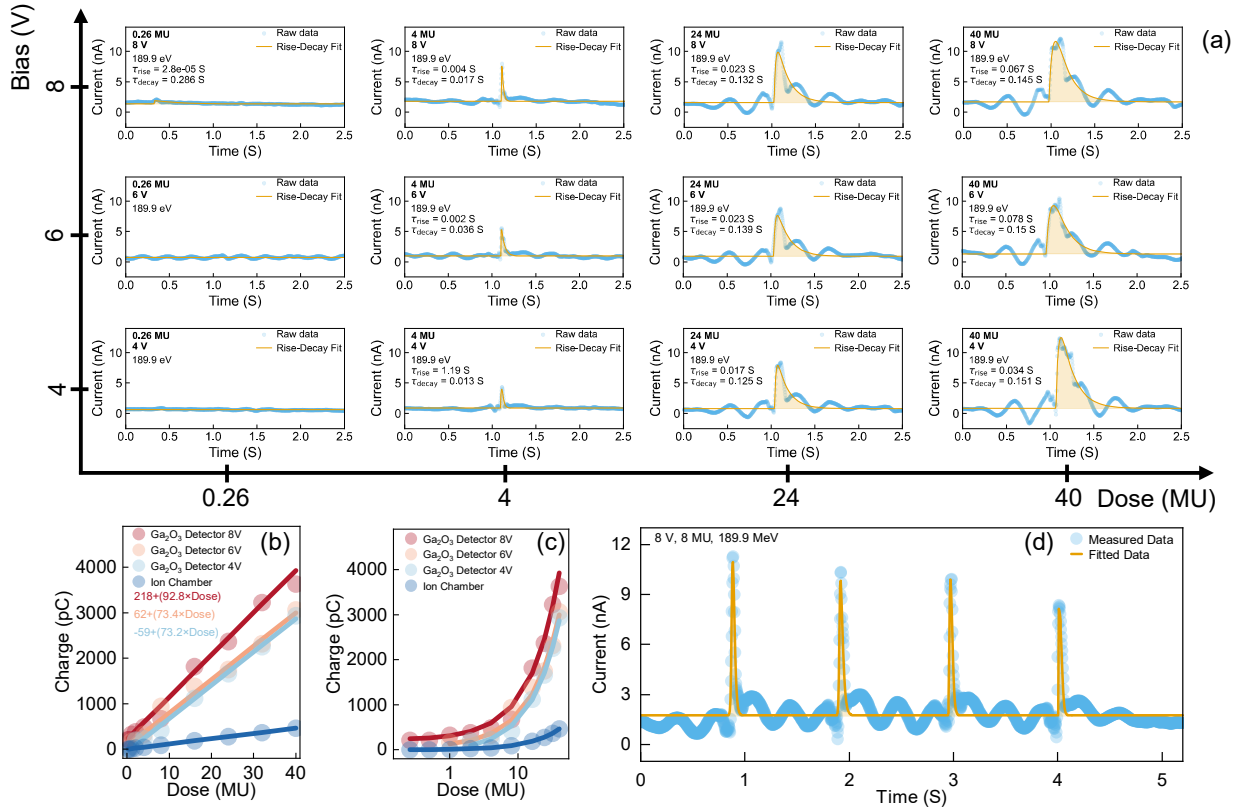

**Fig. 4.** Detection of proton beam in a single-pulse mode. (a) Transient current response of the GOPD biased between 4 V-8 V receiving a total dose of 0.26 MU to 40 MU. Comparison of the total collected charge between the ion chamber and the GOPD biased at different voltages for the single-pulse proton radiation in linear (b) and logarithmic (c) scales. (d) Transient current response of the GOPD biased at 8 V receiving four pulses of proton radiation with 8 MU per pulse. The equivalent dose in Gy for 0.26 MU, 4 MU, 24 MU, and 40 MU is 3.0 mGy, 0.0464 Gy, 0.278 Gy, and 0.4639 Gy, respectively.

The total charge detected by the ion chamber is compared to the GOPD biased at 4 V, 6 V, and 8 V on a linear scale in Fig. 4(b) and a logarithmic scale in Fig. 4(c). The GOPD at all the tested biases demonstrated a linear charge response with respect to dose, which is similar to the ion chamber. The total charge generated by the proton radiation is defined as

$$C = \int_{t_2}^{t_1} I \, dt \qquad (2)$$

where $C$ represents the total charge, $t_1$ is the proton beam start time, and $t_2$ is the turn-off time. The ion chamber and the GOPD exhibited a linear response. As the bias increases, both the slope and y-intercept of the collected charge also increase, with the linear fitting equations for the detector at different biases provided in Fig. 4(b). The increase in slope is attributed to the detector's enhanced responsivity at higher voltages, resulting in a larger current and, consequently, greater collected charge. Meanwhile, the rise in the y-intercept is due to the higher bias inducing an increased leakage current in the signal, leading to an increased DC offset. The sensitivity of the ion chamber is 1.35 nC Gy$^{-1}$, while the sensitivity of the GOPD is calculated to be 10.51 nC Gy$^{-1}$ at 8 V bias, 8.69 nC Gy$^{-1}$ at 6 V, and 8.32 nC Gy$^{-1}$ at 4 V. Figures 4(d) present the current transient response of the detector to four proton pulses at bias voltage of 8 V. While the GOPD clearly distinguished all four pulses, the observed variation in pulse amplitude may be attributed to either inherent fluctuations in beam output or signal distortion caused by cable-induced RC time-constant effects. This is discussed in more detail in the following section.

The GOPD also measured a raster-scanned proton beam, which consists of multiple pulses of a fixed dose with spatial movement in a snaking pattern between each pulse. For this experiment, a 5 cm × 5 cm area was irradiated with spot doses ranging from 0.5 MU to 8 MU, covering a total of 441 spots. The transient response of the detector for the raster-scanned data at varying proton doses is shown in Figs. 5(a)-5(e). These figures clearly demonstrate that both the irradiation duration and the current increase as the dose per spot increases. The extended current duration is attributed to the accelerator's pulsed beam delivery, where higher doses per pulse require more time to fully emit the proton radiation, leading to longer irradiation periods. Additionally, the individual current pulses visible in Figs. 5(a)-5(e) correspond to proton pulses delivered at distinct positions by the accelerator. Variations in pulse current are likely due to the movement of the proton beam, which may alter the amount of proton radiation penetrating the GOPD. The initial current peak is caused by the MEVION S250i accelerator emitting a low-dose pulse to determine if there is any location mismatch that needs to be corrected.

Figure 5(f) shows the total charge detected by the GOPD from raster-scanned proton irradiation, which was calculated using Eq. (2). The GOPD showed a linear relationship between total charge and proton dose. As shown in Fig. 5(f), a linear fit was applied to quantify the relationship between proton dose and total charge.

To assess stability, the total charge measured by the GOPD under varying bias voltages (4 V–8 V) was compared to that obtained from a commercial ion chamber. The total charge collected during each test was calculated using Eq. (2), and the results were normalized to the charge measured in the 1st test of each detector. This normalization facilitated a direct comparison of relative variation between the GOPD and the ion chamber. The resulting normalized charge values for both detectors are presented in Fig. 5(g). At bias voltages of 4 V and 6 V, the GOPD exhibits variations of 1%, which is comparable to those of the ion chamber. However, when the GOPD bias is increased to 8 V, the variation increases to 5%, indicating reduced stability at higher operating voltages. This increased variability—also evident in Fig. 4(d)—may stem from enhanced susceptibility to electronic noise originating from the long signal cables, which is exacerbated at higher biases due to increased gain and higher leakage currents.. These findings underscore the importance of optimizing the applied bias to balance sensitivity and signal stability.. These results highlight the strong potential of the GOPD for accurate dose verification in proton therapy.

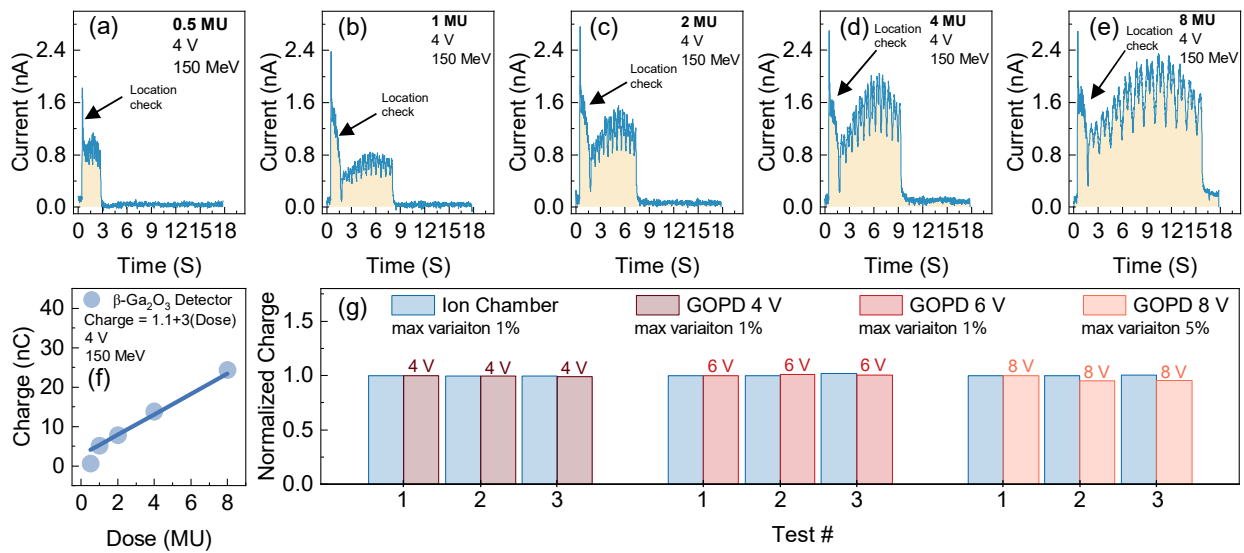

**Fig. 5**. Detection of proton beams with in raster-scan mode. (a–e) Transient current responses of the GOPD biased at 4 V to raster-scanned proton pulses with per-pulse doses of (a) 0.5 MU, (b) 1 MU, (c) 2 MU, (d) 4 MU, and (e) 8 MU. (f) Total collected charge from the GODP at 4 V under raster-scanned proton beam exposure. (g) Comparison of signal variation across repeated measurements: normalized variation for GOPD biased at 4 V, 6 V, and 8 V, versus that of the ionization chamber under the same conditions, with the charge normalized to the first measurement. The equivalent dose in Gy for 0.5 MU, 1 MU, 2 MU, 4 MU, and 8 MU is 0.412 Gy, 0.832 Gy, 1.664 Gy, 3.327 Gy, and 6.744 Gy, respectively.

**Conclusion**

This study demonstrates the feasibility of using a GOPD for proton beam monitoring in IMPT. The detector's performance was systematically evaluated and compared with a commercial ion chamber across a wide range of operating conditions, including single-pulse exposures (0.26 MU to 40 MU), multi-pulse delivery at 8 MU per pulse, and raster-scanned beam profiles with spot doses from 0.5 MU to 8 MU. The GOPD exhibited a linear charge-to-dose response over the entire dose range tested, particularly at a bias voltage of 8 V, enabling detection of even the lowest dose deliverable by the MEVION S250i accelerator. Furthermore, the detector demonstrated stable and repeatable behavior across multiple tests under appropriate biases. These results highlight the strong potential of GOPDs as robust, radiation-hard alternatives to conventional dosimetry tools for real-time, high-resolution dose verification in IMPT.

**AUTHOR DECLARATIONS**

**Conflict of Interest**

The authors have no conflicts to disclose.

**Author Contributions**

Hunter Ellis: Fabrication(lead); Data curation (lead); Formal analysis (lead); Writing-original draft (lead). Imteaz Rahaman: Fabrication (equal). Apostoli Hillas: data curation (equal). Botong Li: review & editing (supporting). Vikren Sarkar: Conceptualization (lead); Data curation (equal);

Resources (lead). Kai-Fu: Conceptualization (equal); Supervision (lead); Project administration (lead); Resources (equal).

## ACKNOWLEDGEMENT

The authors acknowledge the support from the Pilot Funding through the HCI/Engineering Innovation in Cancer Engineering (ICE) Partnership Seed Grants, the University of Utah start-up fund, and PIVOT Energy Accelerator Grant U-7352FuEnergyAccelerator2023. This work was performed in part at the Utah Nanofab Cleanroom sponsored by the John and Marcia Price College of Engineering College of Engineering and the Office of the Vice President for Research. The authors appreciate the support of the staff and facilities that made this work possible.

## DATA AVAILABILITY

The data that support the findings of this study are available from the corresponding authors upon reasonable request.